# Controlled phase quantum gate on off-resonant interaction of photon with three-level atom in single-mode resonator

S.A. Moiseev[1,2] and S.N. Andrianov[1,3]

[1] Kazan Quantum Centre, Kazan National Research University, 420015 Kazan, Russia
[2] Kazan Physical-Technical Institute, Russian Academy of Sciences, 420029 Kazan, Russia
[3] Institute of Perspective Research, Tatarstan Republic Academy of Sciences, 420111 Kazan, Russia.

## Abstract

The optical scheme is proposed for realization of controlled phase quantum gate based on the off-resonant interaction of photon with three-level atom in single mode QED-cavity. Possible physical implementation of this scheme is discussed.

## 1. Introduction

Photon resonators and single atoms are the convinient element base for the construction of quantum information devices [1-5]. Large progress in the elaboration of quantum-nonlinear systems has opened up new horisons for single-photon experiments [6-8] and implementation of deterministic photonic two-qubit gates [3,5,9]. There is great interest to the robust implementation of optical phase shift $\pi$ controlled by a single photon [1]. In the orignial protocol [1], the authors consider a resonant interaction of a photon with an atom in QED cavity which could provide a perfect phase $\pi$-shift for the reflected photon. In comparison with traveling two-photon schemes based on double EIT [10-13], first photon is stored initially on the atomic long-lived states in one side QED cavity, and second photon is launched later to the QED cavity reflecting backward with additional pi-shift after resonant interaction with the atom. Now, the scheme seems to be more realistic in comparison with the proposed travelling schemes [14,15]. Moreover the considered QED-based scheme could provide stronger photon-photon coupling due to the Purcell effect in photon-atom interaction [16].

Recent experiment [3] has been demonstrated successfully the original protocol [1] with a fidelity f>71% by using a single $^{87}$Rb atom trapped in three-dimensional optical lattice. However the achieved basic parameters of the phase gate protocol should be still improved. In this our work, we propose some modification of the original protocol by using off-resonant interaction between a photon and three-level atom in QED cavity and we discuss the potential advantages of this scheme.

## 2. Quantum gate of controlled phase

Scheme for realization of two-photon gate is presented in Fig.1. It includes a single mode optical QED cavity with resonance frequency $\omega_o$ that contains three-level atom with optical transition frequencies ($\omega_{31}$ и $\omega_{32}$) chosen close to the cavity frequency. It is supposed that lower long lived atomic energy levels are separated one of the other at the defined frequency interval $\omega_{21}$. Single photon fields are launched along the cavity axis and supplement intensive control laser field with frequency $\omega_c$ propagates orthogonally to the cavity axis and with single photon field provides the Raman resonance transition of atom between its ground states $|1\rangle$ and $|2\rangle$ (where $\omega_c$-$\omega_o$=$\omega_{21}$). Scheme of the atomic quantum transitions and the interaction of atom with single photon field is depicted in Fig.2. The cavity freqency $\omega_o$ is chosen to be offset relatively to atomic frequency transitions 1-3 and 2-3 at equal frequency intervals: $\omega_{31}-\omega_o = \omega_o-\omega_{32} = \Delta/2$, where $\Delta=\omega_{21}$. For the realization of considering photon-photon interaction, atom is prepared initially in the ground state $|1\rangle_a$. Then, the photon qubit in the state $|\psi\rangle_c = \alpha_c|0\rangle_c + \beta_c|1\rangle_c$ is sending into the cavity along with laser control field (where $|\alpha_c|^2 + |\beta_c|^2 = 1$, $|1\rangle_c = \int d\omega f_c(\omega)\hat{a}^+_\omega|0\rangle$, $|0\rangle$ is vacuum state of light, $f_c(\omega)$ is wave function of photon with spectral width $\delta\omega_c$, $\int d\omega |f_c(\omega)|^2 = 1$). It is well known that it is possible to provide with high probability the transfer of atomic quantum state on the ground states of atom by choosing appropriate parameters of control laser radiation. For example, it is possible with an exponentially rising shape $\tilde{f}_c(t) = -\frac{i}{\sqrt{2\pi}}\int d\omega f_c(\omega)\exp\{-i(\omega-\omega_2)t\} \sim \exp\{t/(2\delta t_c)\}\eta_-(t)$ (where $\eta_-(t)$={1(t<0),0(t>0)} of the photon wavepacket, $\delta t_c$ is temporal duration of the wave packet) [17]. Accounting initial state $|\Psi\rangle_1 = |\psi\rangle_c|1\rangle_a$, we get in this case just after the absorption of single photon field the following state:

$$|\Psi\rangle_2 = \hat{U}|\Psi\rangle_1 = [\alpha_c|1\rangle_a - \beta_c|2\rangle_a]\otimes|0\rangle, \tag{1}$$

where the sign «-» is the result of unitary evolution.

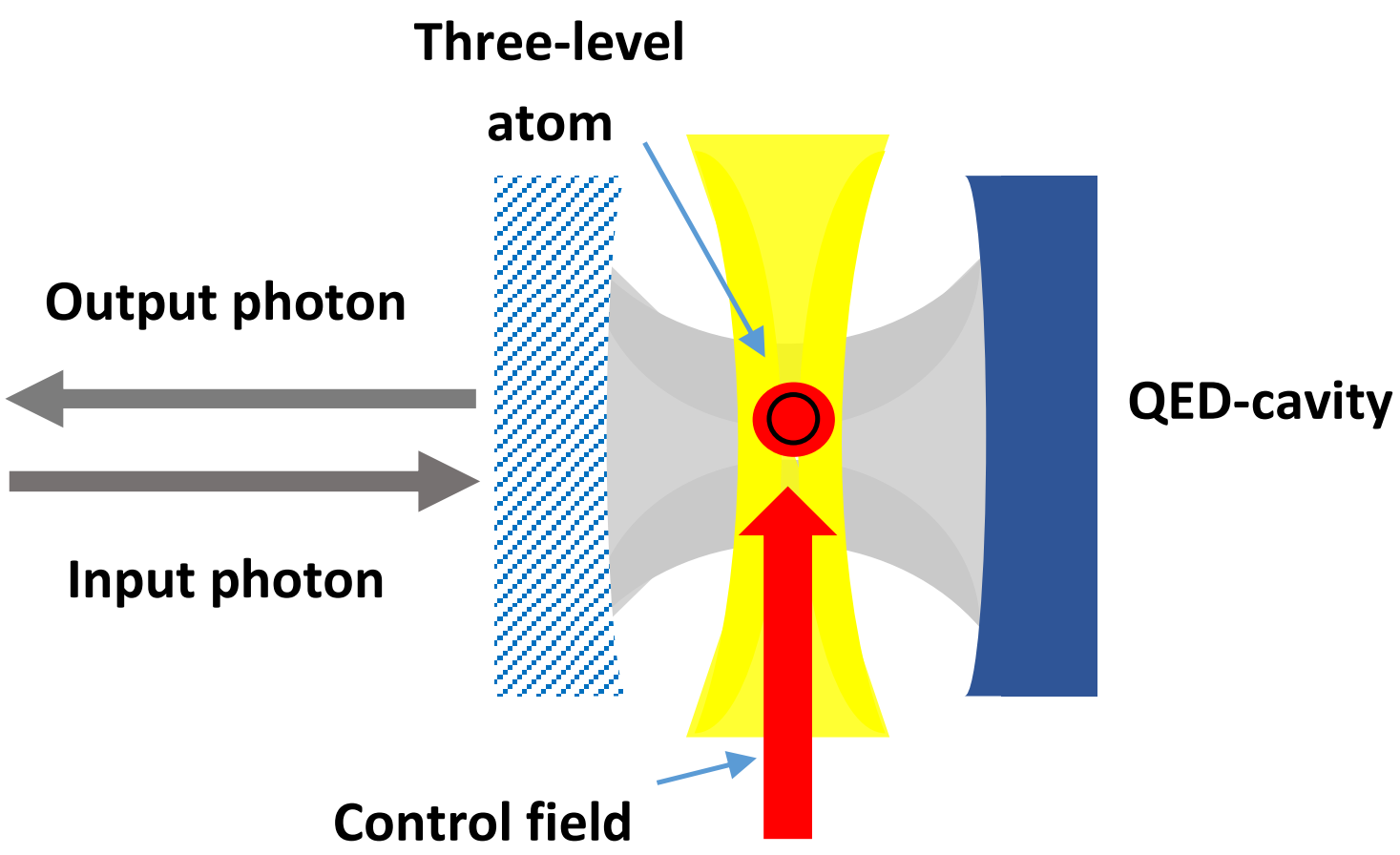

Fig.1. Spatial scheme of atom in optical QED cavity controlled by an additional laser field.

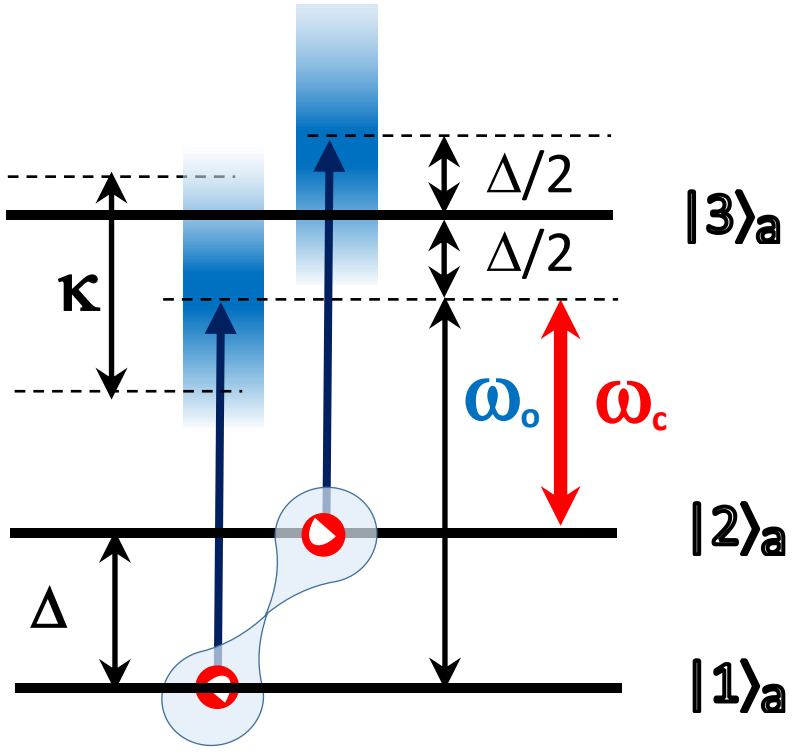

Fig.2. Atomic energy levels and used quantum transitions, $\omega_o$ – frequency of QED cavity, $\omega_c$ – frequency of the control laser field.

Further, target (signal) photonic wave packet is launched in the cavity. Its state we define as $|\psi\rangle_s = |1\rangle_s$, where index «s» in $|1\rangle_s$ denotes a signal photon, that determines the common state of atom and field as follows $|\Psi\rangle_{ph+a} = |1\rangle_s \otimes |\Psi\rangle_2$. The Hamiltonian of complete system has the form: $\hat{H} = \hat{H}_0 + V$, where $\hat{H}_0 = \hat{H}_a + \hat{H}_c + \hat{H}_f$ is main Hamiltonian including atomic Hamiltonian $\hat{H}_a = \hbar\omega_{21}\hat{P}_{22} + \hbar\omega_{31}\hat{P}_{33}$ and Hamiltonian of cavity field mode $\hat{H}_c = \hbar\omega_o\hat{a}^+\hat{a}$, Hamiltonian of free radiation field $H_f = \hbar\int d\omega\omega b^+(\omega)b(\omega)$ and Hamiltonian of interaction

$$V = \hbar g_{fc}\int d\omega\left(b(\omega)\hat{a}^+ + b^+(\omega)\hat{a}\right) + \hbar g[\left((\hat{P}_{13} + \hat{P}_{23})\hat{a}^+ + (\hat{P}_{31} + \hat{P}_{32})\hat{a}\right)]. \tag{2}$$

Supposing that the resonance interaction of light with atom ($\frac{1}{2}|\Delta| << \omega_o$) is dominating, we find the wave function of the studied system as

$$|\Psi(t)\rangle = |\Psi_{a+f}(t)\rangle + |\Psi_{a+c}(t)\rangle + |\Psi_{a+v}(t)\rangle, \tag{3}$$

where $|\Psi_{a+f}(t)\rangle = \alpha_c|1\rangle_a \otimes \int d\omega f_1(\omega,t)b^+(\omega)|0\rangle - \beta_c|2\rangle_a \otimes \int d\omega f_2(\omega,t)b^+(\omega)|0\rangle$ describes free field, with that atom and cavity mode are in the ground states, $|\Psi_{a+c}(t)\rangle = [\chi_1(t)\alpha_c|1\rangle_a - \chi_2(t)\beta_c|2\rangle_a]\otimes\hat{a}^+|0\rangle$ corresponds to the single photon excitation of cavity with atom being in the ground state, $|\Psi_{a+v}(t)\rangle = [\xi_1(t)\alpha_c - \xi_2(t)\beta_c]|3\rangle_a \otimes |0\rangle$ corresponds to excitation of atom in upper optical state. We find unknown functions in states $|\Psi_{a+f}(t)\rangle, |\Psi_{a+c}(t)\rangle, |\Psi_{a+v}(t)\rangle$ using input-output approach to the description of light interaction with atom in the cavity [18] and acounting (see Fig.2) sufficiently narrow spectrum of the input

signal photon in comparison with the frequency offset relatively to the precise resonance $\delta\omega_s << |\Delta|$:

$$\frac{d}{dt}\chi_n = -\frac{\kappa}{2}\chi_n - ig\xi_n + \sqrt{\kappa}\tilde{f}_s, \tag{4}$$

$$\frac{d}{dt}\xi_1 = -\tfrac{1}{2}(i\Delta - i\gamma)\xi_1 - ig\chi_1, \tag{5}$$

$$\frac{d}{dt}\xi_2 = \tfrac{1}{2}i(\Delta + i\gamma)\xi_2 - ig\chi_2, \tag{6}$$

where $\tilde{f}_s(t) = \int d\omega f_s(\omega) e^{-i(\omega-\omega_o)t}$ , $\kappa = 2\pi g_{fc}^2(\omega_o)$ and the decay constant γ is introduced phenomenologically for evaluation of final relations.

Finding the solutions of equations (4), (5) and (6) and using the well-known relations for input and output fields which take the following form for the field wave functions $f_{1,2(in)} + f_{1,2(out)} = \sqrt{\kappa}\chi_{1,2}(t)$, (where $f_{1,2(in)}(\omega) = f_s(\omega)$), we get for the output fields

$$f_{1(out)}(\nu + \omega_o) = \frac{1 + i\frac{2\nu}{\kappa} - i\frac{2g^2}{\kappa(\Delta - \nu - i\gamma)}}{\{1 - i\frac{2\nu}{\kappa} + i\frac{2g^2}{\kappa(\Delta - \nu - i\gamma)}\}} f_s(\omega), \tag{7}$$

$$f_{2(out)}(\nu + \omega_o) = \frac{1 + i\frac{2\nu}{\kappa} + i\frac{2g^2}{\kappa(\Delta + \nu + i\gamma)}}{\{1 - i\frac{2\nu}{\kappa} - i\frac{2g^2}{\kappa(\Delta + \nu + i\gamma)}\}} f_s(\omega). \tag{8}$$

Accounting narrow spectrum of the input signal in (7)-(8): $2|\nu|/\kappa << 1$, $2|\nu|/\Delta << 1$, $\gamma/\Delta << 1$ we get:

$$f_{1(out)}(\nu + \omega_o) \cong \exp\{-i2\varphi\} f_s(\omega), \tag{9}$$

$$f_{2(out)}(\nu + \omega_o) \cong \exp\{i2\varphi\} f_s(\omega), \tag{10}$$

where $tg(\varphi) = 2g^2/\kappa\Delta$.

Choosing interaction parameters that satisfy the condition $2g^2/\kappa\Delta = 1$ ($\varphi = \pi/4$) (we call it the matching condition of phase π gate), we get as a result after the termination of interaction of photon with atom:

$$|\Psi(t >> \delta t)\rangle = \int d\omega f_s(\omega, t) b^+(\omega)|0\rangle \otimes \{\alpha_c \exp\{-\tfrac{1}{2}i\pi\}|1\rangle_a - \beta_c \exp\{\tfrac{1}{2}i\pi\}|2\rangle_a\}. \tag{11}$$

At the last stage, we direct the reading laser radiation to the cavity atom analogously to the first stage of control photon recording that leads to reradiation of this photon in the state $|1\rangle_c$ for the term of wave function with atom in the state $|2\rangle_a$, leading to the following wave function:

$$\left|\Psi_{out}(t)\right\rangle = \exp\{-\tfrac{1}{2}i\pi\}\{\alpha_c\left|0\right\rangle \otimes \left|1\right\rangle_s - \beta_c\left|1\right\rangle_c \otimes \left|1\right\rangle_s\} \otimes \left|1\right\rangle_a . \qquad (12)$$

Solution (12) shows that the interaction of photons in the cavity leads to the supplement phase shift equal π that corresponds to the realization of control phase gate.

## 3. Conclusion

Thus, we have shown the feasibility of the photon-photon controlled π phase gate due to the subsequent off resonant interaction of two photons with three-level atom located in a single-mode QED-cavity. Herein, we have considered the case of large photon-atom frequency offset. Therefore, it deals not with absorption and re-radiation of real photon but rather with its off-resonant scattering. Small spectral detuning from exact resonance is determined by the spectral matching condition $\Delta = 2g^2/\kappa$. For experimental realization of this condition, it is preferable to use atoms (natural or artificial) with large enough constant communication "g", for example, it could be quantum dots [19-21], atoms of rubidium [22], NV center (and other new centers) in diamond [23]. At the same time, to increase the "g" photon-atom coupling constant, it is possible to use nanowaveguide resonators and photonic crystal resonators actively developed in recent decade [24-26]. It is worth noting that the analyzed scheme can be even easily realized in microwave quantum circuits where strong coupling constants “g” between three-level superconducting system and microwave photon is achievable [27].

Though atom-field coupling is weaker in considered case than in the proposal [1] based on the resonant photon-atom interaction, the influence of decoherence effects can be made weaker in the considered scheme. It paves the way for possible applications in quantum informatics. The proposed photon-control phase gate could be useful for the experimental implementation of basic photon-photon gates in QED-cavity, which are required for implementation of various photon-processing protocols in quantum communication [28], for the preparation of photonic cluster states [29], creation of quantum RAM schemes [30]. Also, it can be used for the construction of deterministic optical quantum computer.

### Acknowledgements

The research has been supported by the Russian Science Foundation through the Grant No. 14-12-01333-P.